\documentclass[preprint]{elsarticle}
\usepackage[utf8]{inputenc}
\usepackage{amsmath}
\usepackage{amssymb}
\usepackage[normalem]{ulem}
\usepackage{gensymb}
\usepackage{graphicx}
\usepackage[colorinlistoftodos]{todonotes}
\usepackage{lineno}
\usepackage{hyperref}
\hypersetup{colorlinks,allcolors=blue}
\usepackage{siunitx}
\usepackage{natbib}
\usepackage{float}
\usepackage{appendix}
\usepackage{caption}
\usepackage{subcaption}
\usepackage[a4paper]{geometry}
\begin{document}

\title{Ultraviolet-induced fluorescence  of poly(methyl methacrylate) compared to 1,1,4,4-tetraphenyl-1,3-butadiene down to 4 K}

\author[a,b]{E.~Ellingwood}
\author[a]{H.~Benmansour}
\author[a]{Q.~Hars}
\author[a]{J.~Hucker}
\author[a]{V.~Pereimak}
\author[a]{J.~M.~Corning}
\author[a]{P.~Perrin}
\author[c]{G.~R.~Araujo}
\author[a]{P.C.F.~Di~Stefano}
\ead{distefan@queensu.ca}
\author[d,e,b]{M.~Ku\'zniak}
\author[f,c]{T.R.~Pollmann}
\author[g]{M.~Hamel}
\author[e]{M.G.~Boulay}
\author[e]{B.~Cai\fnref{fn1}}
\author[e]{D.~Gallacher\fnref{fn2}}
\author[a,b]{A.~Kemp}
\author[e]{J.~Mason}
\author[a,b]{P.~Skensved}
\author[a,b]{M.~Stringer}

\address[a]{Department of Physics, Engineering Physics \& Astronomy, Queen’s University, Kingston, ON, K7L 3N6, Canada}
\address[b]{Arthur B. McDonald Canadian Astroparticle Physics Research Institute, Queen’s University, Kingston ON K7L 3N6, Canada}
\address[c]{Department of Physics, Technische Universit\"at M\"unchen, 80333 Munich, Germany}
\address[d]{AstroCeNT, Nicolaus Copernicus Astronomical Center, Polish Academy of Sciences, Rektorska 4, 00-614 Warsaw, Poland}
\address[e]{Department of Physics, Carleton University, Ottawa, K1S 5B6, ON, Canada}

\address[f]{Nikhef and the University of Amsterdam, Science Park, 1098 XG Amsterdam, The Netherlands}
\address[g]{Université Paris-Saclay, CEA, List, F-91120 Palaiseau, France}

\fntext[fn1]{Currently located at Research Services, Queen's University, Kingston, ON, K7L 3N6, Canada}
\fntext[fn2]{Currently located at the Department of Physics, McGill University, Montreal, QC, H3A 2T8, Canada}



\begin{abstract}
    Several particle-physics experiments use poly(methyl methacrylate) (a.k.a. PMMA or acrylic) vessels to contain liquid scintillators. Superluminal charged particles emitted from radioactive impurities in or near the acrylic can emit Cherenkov radiation in the ultraviolet (UV) spectra range.  If acrylic fluoresces in the visible range due to this UV light, it could be a  source of background in experiments where the main signal is visible scintillation light, or   UV scintillation light that is absorbed and re-emitted at visible wavelengths by a wavelength shifter.  Some of these experiments operate at low temperature.
    The fluorescence of these materials could change with temperature so we have studied the fluorescence of the acrylic  used in the DEAP-3600 experiment down to a temperature of 4~K, and  compared it to the common wavelength shifter 1,1,4,4-tetraphenyl-1,3-butadiene (TPB). The light yield and wavelength spectra of these materials were characterized by exciting the sample with 285~nm UV light which acted as a proxy for Cherenkov light in the detector.
    Spectral measurements indicate at least part of the fluorescence of the acrylic is due to additives.  Time-resolved measurements show the light yields of our acrylic sample, TPB sample, and the relative light between both samples, all increase when cooling down. At room temperature, the light yield of our acrylic sample relative to the TPB sample is 0.3~\%, while it reaches 0.5~\% at 4~K.  The main fluorescence time constant of the acrylic  is less than a few nanoseconds.
\end{abstract}

\date{\today}

\maketitle

\section{Introduction}

Liquid scintillators are used as the detection medium by current and planned particle physics detectors for rare event searches like neutrino and dark matter experiments, such as MicrobooNE~\cite{microboone_collaboration_design_2017}, Daya Bay~\cite{krohn_long-term_2012}, DarkSide~\cite{aalseth_darkside-20k_2017}, and SNO+~\cite{flaminio_current_2016}. Acrylic, which is composed of poly(methyl methacrylate) (PMMA) possibly with trace amounts of additives, is a popular structural material because it is optically transparent and radio-pure.
The DEAP-3600~\cite{DEAP} dark matter experiment uses liquid argon (LAr) as the scintillator in a vessel made of acrylic. The inside of the vessel is coated with 1,1,4,4-tetraphenyl-1,3-butadiene (TPB), a wavelength shifter, which converts 128~nm ultraviolet (UV) LAr scintillation  to visible light that better matches the quantum efficiency of the standard photomultiplier tubes (PMTs) used. However, UV scintillation light may reach the acrylic in uncoated areas.  In addition, fast charged particles can produce Cherenkov UV light within, or entering, the acrylic. For both these mechanisms, the acrylic absorbs UV light and prevents it from reaching the PMTs. 
However, UV-induced fluorescence has been observed in certain types of acrylic~\cite{Tarasenko}; this could contribute to the background in very sensitive rare-event searches.

The wavelength shifting properties of TPB have been investigated previously. Upon UV irradiation, a sample of TPB will emit visible ($\sim 420$~nm) fluorescence corresponding to the $\sim$~nanosecond de-excitation of occupied singlet states~\cite{flournoy_substituted_1994}. It is possible to evaporate TPB onto glass substrates~\cite{Burton}, reflectors~\cite{araujo_rd_2021}, and the inside of large acrylic vessels~\cite{DEAP}. Features of the re-emission peaks and quantum yield can be manipulated by changing various parameters of the sample such as the thickness~\cite{gehman_fluorescence_2011}, incident wavelength~\cite{gehman_fluorescence_2011}, and sample age~\cite{Graybill}.
At $\sim$10~K, the light yield of TPB is $1.3 \pm 0.1$ times higher than at room temperature~\cite{Francini}.  
While studies of TPB at vacuum ultraviolet (VUV) wavelengths like 128~nm are rare, longer-wavelength UV irradiation experiments can provide a basis for extrapolation to the VUV spectrum~\cite{TPB1}.

Investigations of acrylic luminescent properties have been previously carried out~\cite{Piruska,miyashita_autofluorescence_2008}. These are not necessarily representative of all types of acrylic, since impurities, additives, defects, and the surface finish of the material  potentially  contribute to the luminescence signal. Fluorescence and optical properties of acrylic were measured under visible laser light excitation~\cite{Piruska}. Additionally,   the luminescence of laser-etched acrylic chips was found to increase compared to a bulk sample of acrylic~\cite{miyashita_autofluorescence_2008}. Both samples exhibited re-emission at a wavelength of approximately $625$~nm.
Luminescence of nominally pure PMMA was studied in the context of embedded metallic clusters~\cite{molard_pmma}, showing a broad spectrum around 440~nm when excited at 355~nm.
In addition, the luminescence of acrylic was studied under excitation with an electron beam and $\sim 222$~nm UV light~\cite{Tarasenko}. Samples that are 6 and 10~mm thick show a broad band at $\sim 490$~nm. The 3~mm thick sample showed an additional band at 400~nm and the absorption spectrum was shifted from 300~nm to 350~nm, which can be explained by the presence of impurities~\cite{Tarasenko}.

Moreover, investigations were done with acrylic from the same batch that was used for the DEAP-3600 vessel, referred to as AVA for this study. 
At room temperature, the AVA time-resolved response was measured under UV excitations ranging from 130~nm to 250~nm~\cite{RPT_LY}, which showed that the fluorescence of the AVA sample was at most 0.2\% relative to TPB.
In addition, our group has previously measured the AVA and TPB fluorescence spectra under 280~nm excitation in the 300~K to 4~K temperature range~\cite{LIDINE} with samples which will be referred to as AVA3 and TPB3. Fluorescence was clearly visible around 400~nm, and there was a qualitative increase in the light yield during cooling of the sample.
 In this work, we investigate fluorescence of a newer set of AVA and TPB samples that are referred to as AVA1 and TPB1. The samples used in the spectrometer, AVA2 and TPB2, are identical samples to AVA1 and TPB1 used for the time-resolved measurements. We quantify the time-resolved light yield of the AVA1 and TPB1 samples over a range of temperatures, including 87~K and 4~K, the boiling points of argon and helium.  Compared to earlier work~\cite{LIDINE}, we have reduced systematic uncertainties related to the stability of the excitation, which should improve the precision of our light yield measurements.

\section{Sample and Equipment Details}

\begin{figure}[H]
        \centering
        \includegraphics[width=0.6\textwidth]{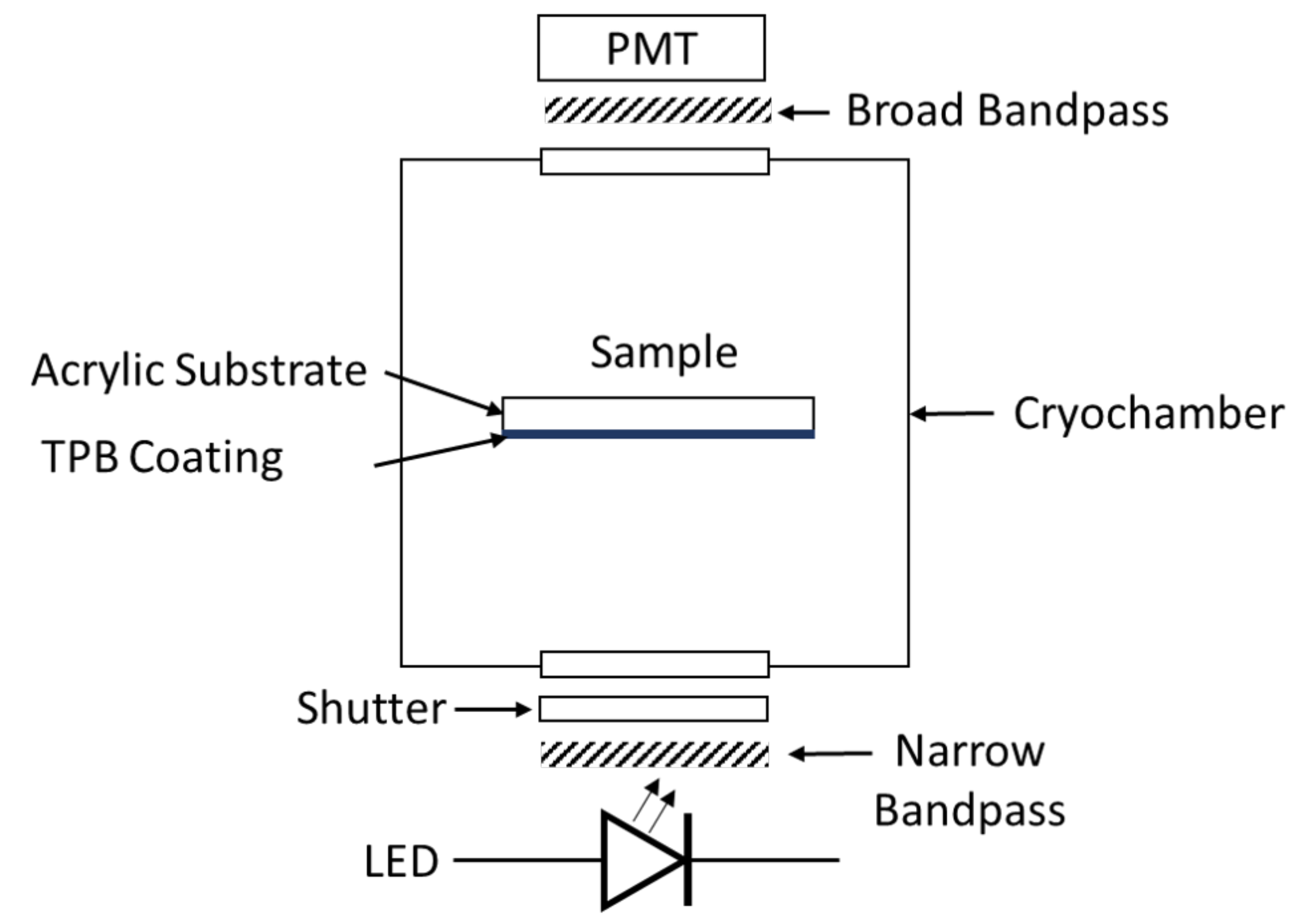}
        \caption{Optical cryostat sample setup for time-resolved measurements of the sample fluorescence light.}
        \label{fig:setup}
\end{figure}

The purpose of this experiment was to measure how the relative light yield of the AVA compared to TPB changes with temperature under 285~nm UV excitation. The measurements were done in a closed-cycle optical cryostat at the following temperatures: 300, 292, 273, 250, 210, 163, 120, 100, 87, 77, 50, 27, 15, 10, 8, 6, 5 and 4~K. The cryostat has a compact geometry that optimizes the light collection efficiency~\cite{cryostat}. The general setup of the experiment is illustrated in Figure~\ref{fig:setup}.

The samples used for our main study were designed to be attached directly to the cryostat coldfinger and fit within the cryostat.
They are rectangular parallelepipeds, with a thickness in the optical direction of 5~mm, a width of 24~mm, a height of 30~mm, and tabs for fixing to the coldfinger (see Fig.~2 of Ref.~\cite{benmansour_fluorescence_2021}).
Both of the samples used were made from acrylic specifically using the same batch manufactured by Reynolds Polymer Technologies that was used in building the acrylic sphere for DEAP-3600. The faces of the samples were finished to rough sanded quality with Klingspor PS11A P400 sandpaper (42~$\mu$m size). One of the samples was coated on one face with 1~$\mu$m of TPB, the WLS coating that is on the inner surface of the DEAP-3600 acrylic vessel, using thermal vacuum evaporation. 

The setup consists of a $\sim$10~ns FWHM pulse of 285~nm UV light from an LED which passes through a shutter and the windows of each of the three layers of the cryostat into the main chamber. A function generator is used to trigger the Kapustinsky~\cite{Kapustinsky} circuit producing the short pulse fed to the LED, and the same generator pulse acts as the trigger for the data acquisition (DAQ). The light interacts with the sample producing fluorescent light at longer wavelengths than the initial 285~nm UV LED light. This light then passes through a broad bandpass filter with a lower limit of 375~nm which is intended to eliminate any stray UV light from the LED. For the time-resolved measurements, this light is then detected by a Hamamatsu R6095-100 PMT with a super bialkali photocathode to look at the fluorescence pulses and determine the light yield. For the spectrometer measurements, the Horiba spectrometer replaces the PMT in the setup, the broad bandpass filter is removed and the LED is run in DC mode so it is on continuously. This setup is intended to show wavelength dependent features of the samples' fluorescence spectra. 
Spectra were taken before instabilities in the LED were resolved,  meaning that though the shapes of the spectra can be compared, it is more difficult for their intensities. Instabilities were resolved for the time-resolved measurements, as described in a more complete explanation of the system~\cite{benmansour_fluorescence_2021}.

For the time-resolved measurements, the PMT output is fed into one of the digitizer channels of the DAQ system and it is triggered by the function generator pulse which also controls the LED pulsing. 
The DAQ setup consists of a National Instruments PXIe-5160 digitizer which is controlled by a program where all of the relevant parameters can be set. A single event consists of 12~$\mu$s long waveform made up of 30,000 digitizer samples with 0.4~ns between each sample. The first 10\% of the waveform is the pretrigger.
A full data set at a single temperature is 45,000 of these events. 
The time-resolved measurements for the AVA1 acrylic were taken with a $\pm0.1$~V vertical range on the digitizer since the majority of the waveforms show single photoelectron response with small amplitudes. Meanwhile, the more fluorescent TPB1 sample waveforms were measured with a $\pm$2.5~V vertical range to  capture the entire waveform without saturation. 
Lastly, for the spectrometer measurements, the signal is read out in the computer by a program specific to the spectrometer for a 10~s exposure.

In addition to the main samples described above, for comparison, the spectra of three other samples were measured at room temperature only in a  Photon Technology International  (PTI) QuantaMaster fluorescence spectrometer.  These samples  were taken from the acrylic used in the light-guides found in DEAP (DEAP-LG), the DEAP acrylic vessel (AVA4), and the SNO acrylic  vessel (SNO-AV).  The samples can be differentiated by their manufacturer and the additives used to change the optical properties of the respective acrylic samples.  The ultraviolet absorbing (UVA) acrylic used in the light guides of the DEAP experiment was supplied by Spartech,  and the ultraviolet transmitting (UVT) acrylic was used in SNO-AV. 
Though the setups and geometries differ, it should be possible to compare the shapes of the spectra with our main measurements.

\section{\label{sec1}Analysis}

\subsection{\label{sec:PulseAnalysis}Analyzing Time-Resolved Fluorescent Light Pulses}

At each temperature, the 45,000 recorded waveforms are analyzed separately in order to determine the mean light yield from the sample.
There are two initial steps to processing the individual pulse data: baseline subtraction and noise subtraction.
First, the baseline is taken from the average of the first 6\% of the individual waveform so subtracting the data from this baseline produces waveforms with a new baseline around zero. 
\begin{figure}[!h]
    \centering
        \includegraphics[width=0.5\textwidth]{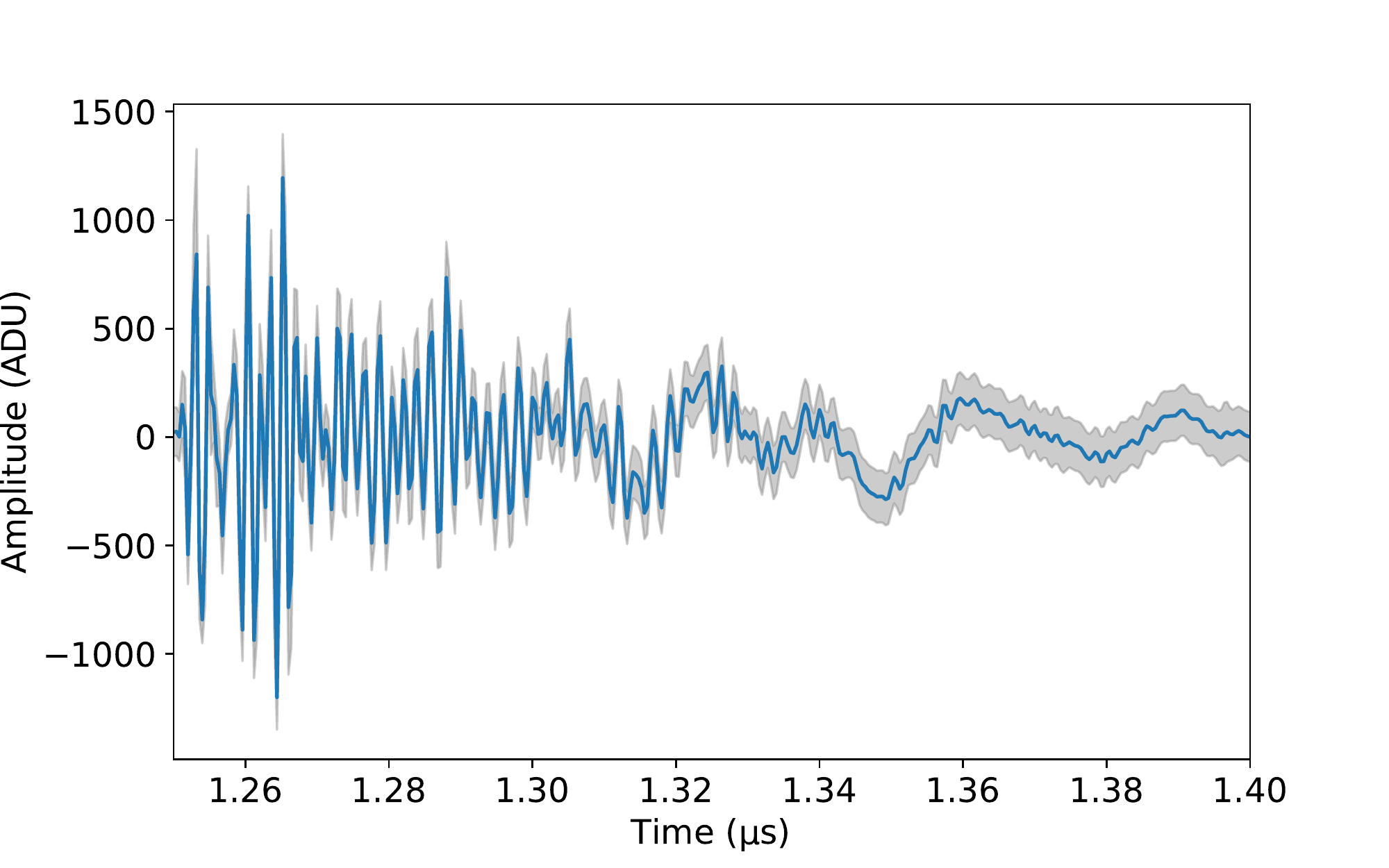}
        \caption{Average signal from all events in a noise run. The grey region is the standard deviation of the distribution of amplitudes at each sample for the 45,000 events in the run.}
        \label{fig:noise_event}
\end{figure}
The next step is to address the noise in the waveforms.
Oscillations around the baseline are observed in the PMT readout from the LED pulses in addition to regular electronic noise. These fluctuations begin in the pretrigger area and overlay on the light pulse in the integration window. To understand this noise, dedicated noise runs are taken at select temperatures by following the same data taking procedure as the pulsed LED data runs except that the shutter between LED and the cryochamber is closed so that no LED light reaches the sample or PMT. Fig.~\ref{fig:noise_event} shows the average  noise-only event. The noise has the same time structure in all events, and can be subtracted from the individual pulses.

\begin{figure}[!h]
    \centering
        \includegraphics[width=0.55\textwidth]{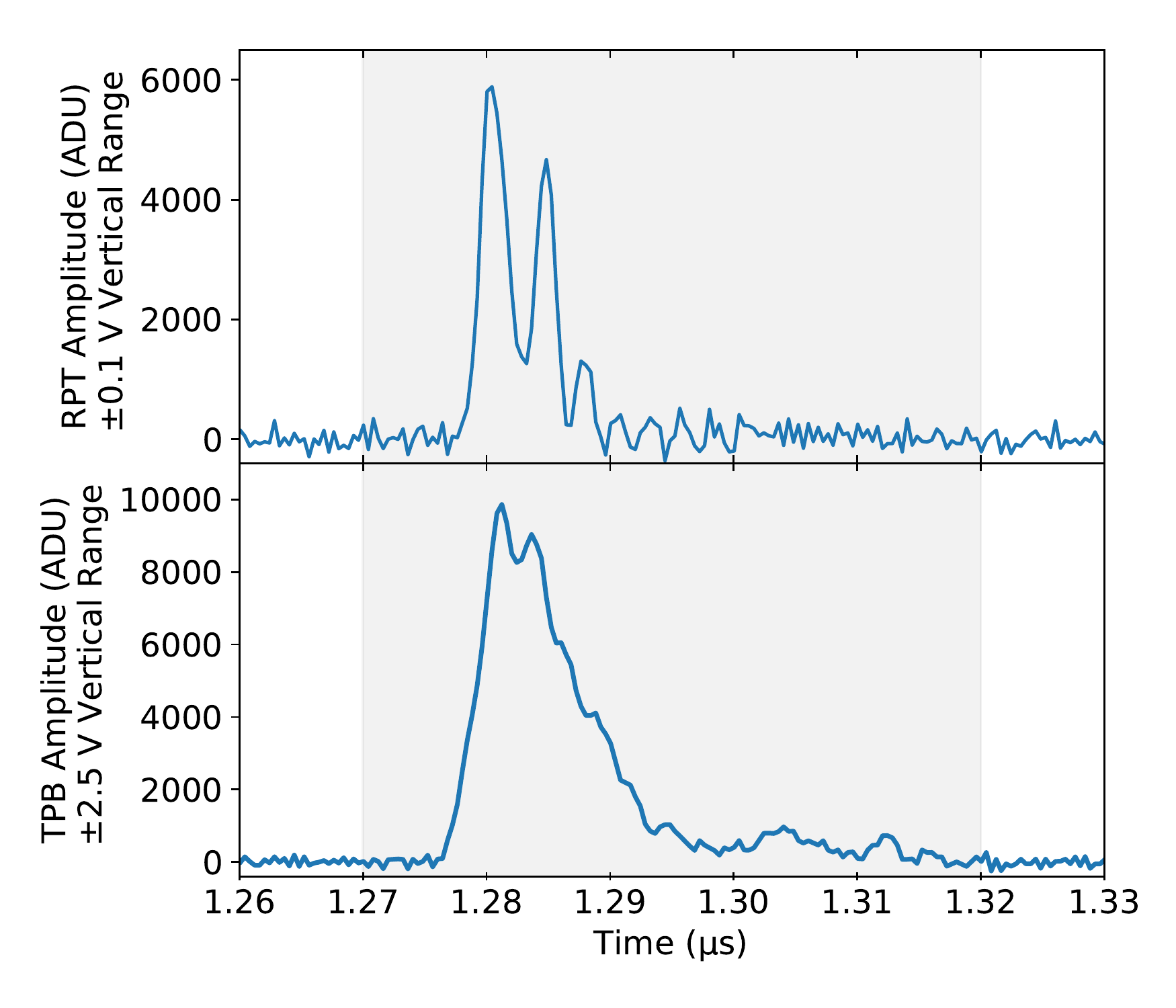}
        \caption{\label{fig:individual_pulses}Examples of individual pulses from the AVA1 and TPB1 samples which are both baseline and noise subtracted. The AVA1 pulse (top) was taken with a $\pm$0.1~V DAQ vertical range to see more detail on a small pulse while the TPB1 data was taken with a $\pm$2.5~V vertical range.  Shaded area is the integration region.}
\end{figure}

Fig.~\ref{fig:individual_pulses} shows example of individual waveforms from AVA1 and TPB1 after baseline and noise were subtracted.  The next step in the analysis is to integrate the pulses over a 50~ns window as illustrated.  This window is set to limit the amount of remaining noise while still containing almost all of the fluorescent light from the sample (also see average pulse shapes in Fig.~\ref{fig:normalized_pulses}).  The distribution of these integrals is then used to determine the light yield, as discussed in the next section.
The integral values of the noise fluctuations are centred on zero, and are negligible compared to those of a single photoelectron (Fig.~\ref{fig:AVA_300K}).

\subsection{\label{sec:LY}Calculating Light Yield}

\begin{figure}[H]
    \centering
    \includegraphics[width=0.7\textwidth]{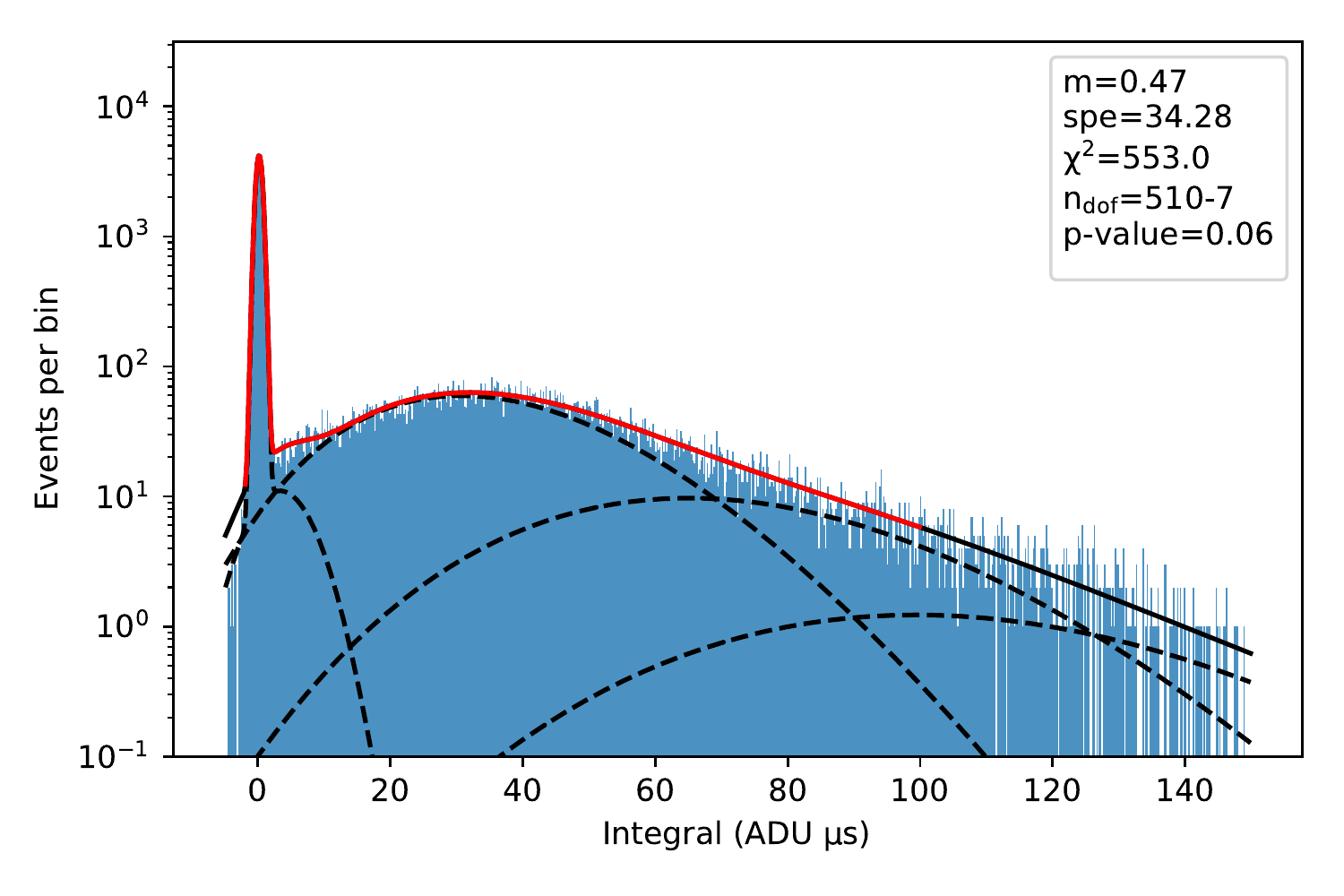}
    \caption{\label{fig:AVA_300K}Integral distribution of AVA1 luminescence at 300~K.  Red curve is overall fit to model in fit range, solid black is overall model outside of fit range, dashed black curves are individual model components.  Parameter $m$ is the average number of photoelectrons per pulse; $spe$ is the average integral of a single photoelectron.  The p-value is the probability of obtaining a larger $\chi^2$ by statistical fluctuations if the model is  correct.  In addition to providing the light yield for the acrylic samples ($m$), analyses of this type applied to acrylic or TPB provide the conversion factor from integrals to photoelectrons for the TPB samples ($spe$).}
\end{figure}

The distribution of pulse integrals obtained in the previous section are used to determine the light yield.
Acrylic is expected to have little to no fluorescence.  Despite having increased the LED intensity for AVA1 compared to TPB1, the PMT records at most a few photoelectrons per trigger.
The integral distribution from these single-photoelectron (SPE) pulses can be fit with a model of the PMT response~\cite{CHIRIKOVZORIN2001310}, as seen in Fig.~\ref{fig:AVA_300K}. The model accounts for the distribution of the number of photoelectrons and provides the average number of photoelectrons and the average single-photoelectron integral. 

In general, TPB is expected to produce much more fluorescent light than acrylic so it is impractical to use the previous model to obtain the light yield.  
Instead, the integral distribution is fit with a skew normal; the mean of that fit is then divided by the average SPE integral determined on the appropriate vertical range to compute  the light yield. 
To avoid saturation while maximizing light output, all of the TPB1 data were taken with the same LED voltage of 13.4~V at all temperatures. The AVA1 data were taken at a higher LED voltage of 13.7~V to ensure that enough light was produced to observe at least a single-photoelectron response. 
To compare the light yields of both materials, the data must be taken with the same excitation intensity (i.e. LED voltage) at a  given temperature. Alternatively, when working with different excitations for each material as in our case, a correction factor, which is the ratio of light yields of one of the materials under both excitations, must be determined.  \ref{sec:CorFact}  details  the correction process.

\section{Results}
\subsection{\label{sec:spectra}Emission Spectra}

The spectra of two samples, AVA2 and TPB2, were taken using the optical cryostat setup and the spectrometer described in~\cite{benmansour_fluorescence_2021} with a 10~s exposure and 285~nm excitation.
The samples used for the spectrometer measurements, AVA2, and TPB2, are made with the same materials, geometry, and have identical properties to AVA1 and TPB1 respectively, which were used in the time-resolved measurements. The spectra observed for the TPB2 and AVA2 samples depend on the properties of the physical sample like the thickness of the acrylic sample, the surface finish, and the thickness of the TPB coating. 

\begin{figure}[H]
    \centering
    \includegraphics[trim={0 50pt 0 80pt},clip,width=0.85\textwidth]{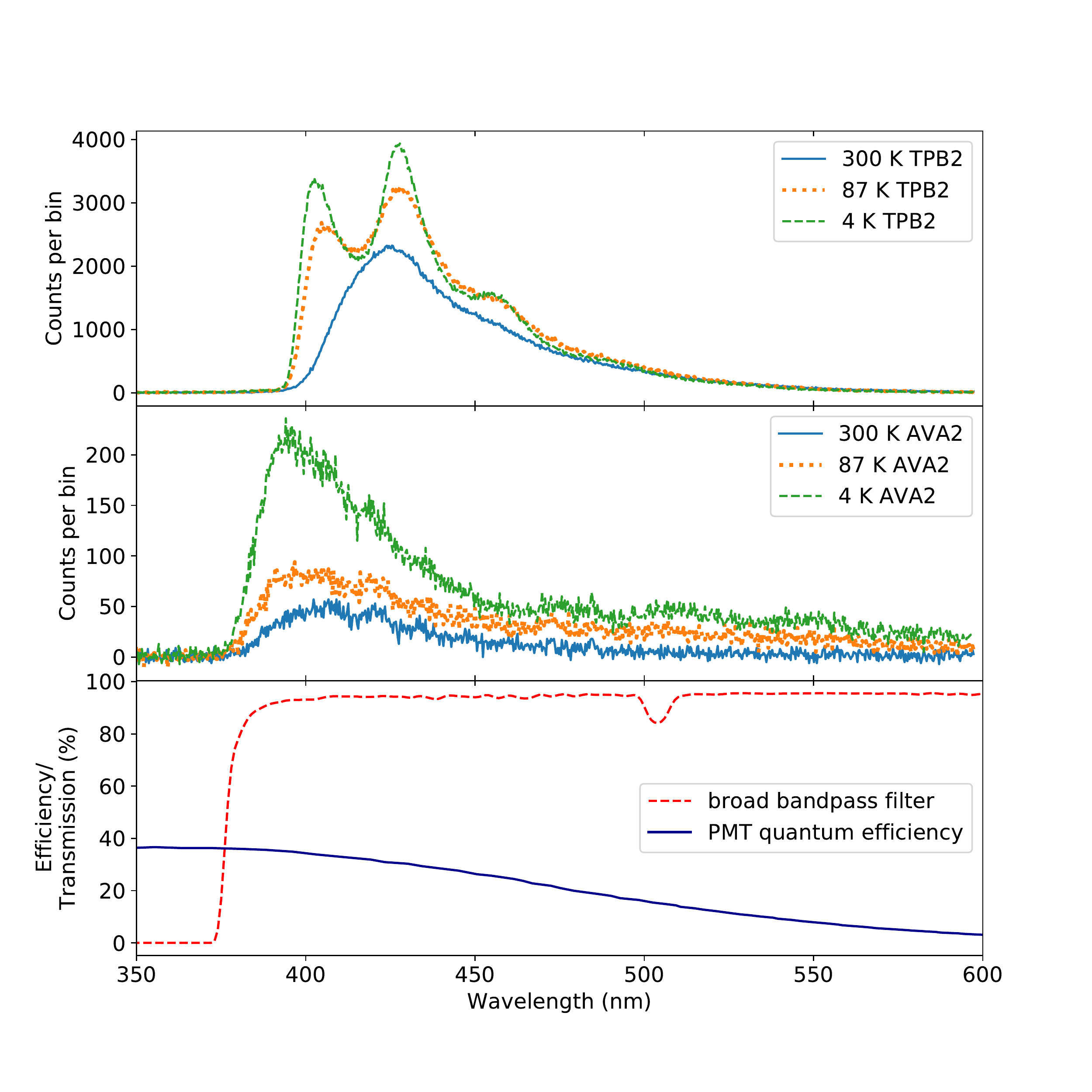}
    \caption{\label{fig:combined_spectra}Wavelength spectra of TPB2 (top) and AVA2 acrylic (middle) luminescence at different temperatures. See text for details.  The bottom plot shows the quantum efficiency of the PMT used in the time-resolved measurements and the broad bandpass filter that was present only for the time-resolved measurements.}
\end{figure}

Figure~\ref{fig:combined_spectra} shows the spectra of the TPB2 sample at 300~K, 87~K and 4~K. It is apparent from the spectra that as the temperature decreases, the overall integral under these spectral curves, and therefore the light yield, increases. In addition, at 300~K there is a single major peak in the spectrum at 425~nm. Heat in the sample allows for the TPB molecules to vibrate, which causes the re-emitted photon to be observed with a Doppler shift related to the temperature of the sample. This Doppler shift causes the luminescent spectral lines to expand, a process known as thermal broadening~\cite{2020Thermal}.  As the temperature diminishes, so does the effect of thermal broadening, and the most prominent peak appears to shift closer to 427~nm. A second peak corresponding to an energy sublevel becomes more prominent at 403~nm, while a third  appears around 455~nm.  

The AVA2 spectrum exhibits a broad main peak around 395~nm, which shifts to slightly longer wavelengths with increasing temperature, possibly cut at lower wavelengths by the broad bandpass filter or self-absorption.  At 4~K, small peaks are observed near 475, 505 and 545~nm.
A similar triple peak structure coming from low-temperature phosphorescence is typical for hydroxyphenyl benzotriazoles~\cite{fluegge_probing_2007,BEAVAN1974925}, some of which are compounds commonly added to polymers to mitigate their degradation by absorbing UVs~\cite{KIRKBRIGHT1970237, neidlinger_polymer_1986, shlyapintokh_effect_1974}. 

In addition, Fig.~\ref{fig:combined_spectra} shows the quantum efficiency of the PMT and the transmission of the broad bandpass filters used in the time-resolved measurements of TPB1 and AVA1 (Sec.~\ref{sec:TPB_LY} and \ref{sec:AVA_LY}). Neither of those components is present in the spectrometer measurements. The inclusion of these curves illustrates how the light yield of the time-resolved measurements could be affected by different efficiencies and transmissions depending on how the spectrum changed with temperature.

\begin{figure}[H]
    \centering
    \includegraphics[trim={0 50pt 0 80pt},clip,width=0.95\textwidth]{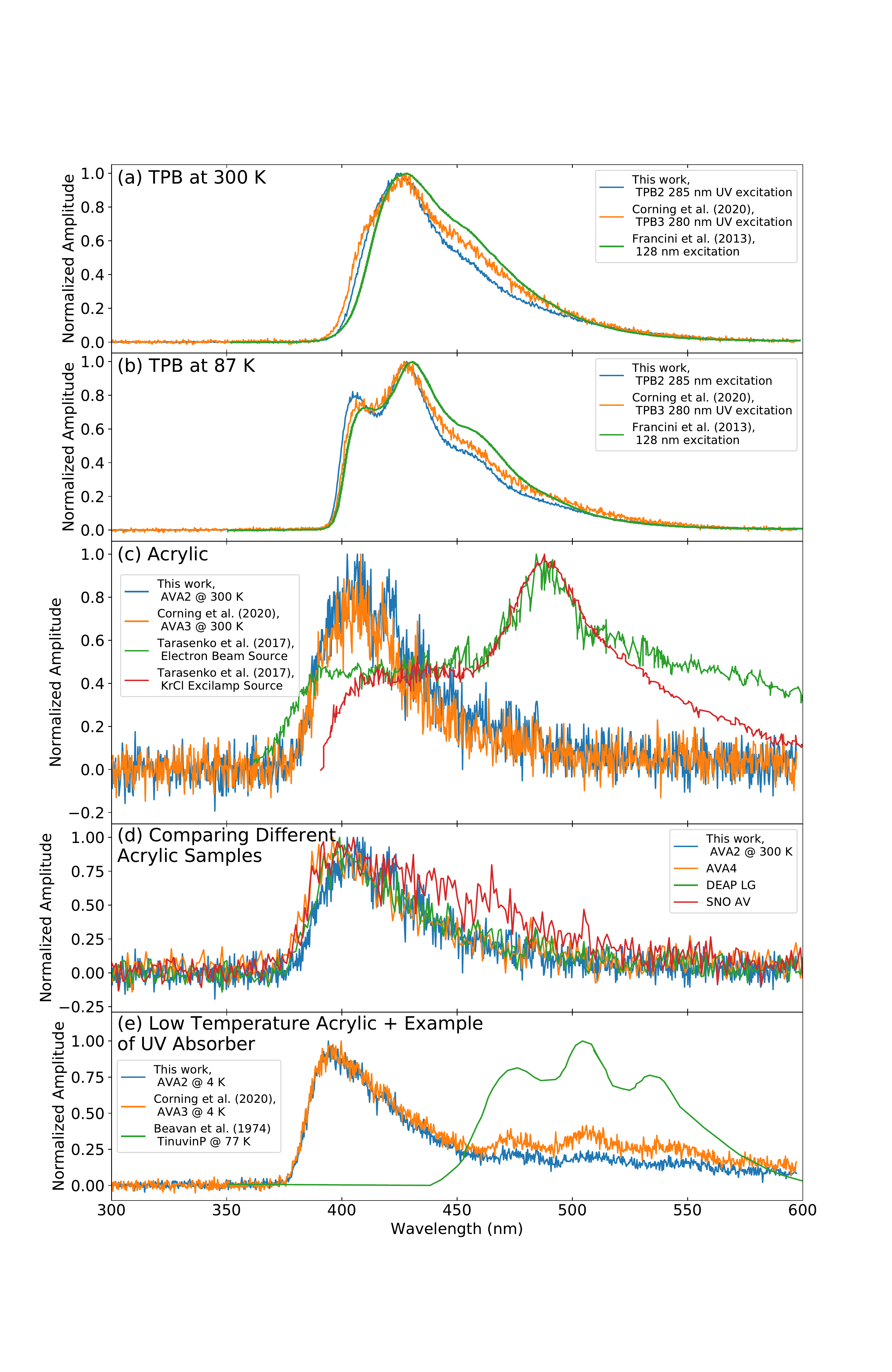}
    \caption{\label{fig:compared_spectra}Comparing wavelength spectra from this work to spectra from literature or previous measurements. (a) spectra of TPB taken at 300~K. (b) spectra of TPB at 87~K. (c) spectra of acrylic at around 300~K. (d) comparing spectrometer measurements of different acrylic samples; the blue and orange curves are data from the spectrometer for this study. (e) sample of Tinuvin~P with EP glass excited at 380~nm, as an example of a compound from the hydroxyphenyl benzotriazole class, which shares similar spectral features. }
\end{figure}

Figure~\ref{fig:compared_spectra} consists of subplots to compare the spectra from this study to previous measurements and literature. Fig.~\ref{fig:compared_spectra}(a) and (b) show that our TPB2 re-emission spectra are consistent with previous work under UV excitation which was done by our group with the TPB3 sample~\cite{LIDINE}, and by other groups at various temperatures with samples of a different origin~\cite{Francini}.  
At room temperature (Fig.~\ref{fig:compared_spectra}(c)), our AVA2 spectra 
are consistent with our previous measurements of material from the same batch (AVA3)~\cite{LIDINE}.
They are also consistent with the spectrum of another sample of the same AVA acrylic from the DEAP-3600 vessel (AVA4), the acrylic used for the light guides in DEAP-3600 (DEAP LG), and with the spectrum of the acrylic used for the SNO/SNO+ vessel (SNO AV), both excited at 280~nm (Fig.~\ref{fig:compared_spectra}(d)).  Elsewhere, nominally pure PMMA excited at 355~nm  shows a broad spectrum  shifted up to 440~nm~\cite{molard_pmma}.
Similar spectra, with some dependence on the excitation wavelength, have been reported for pure copolymers of methyl methacrylate and acrylic acid~\cite{SOSAFONSECA2001327}.
Our spectra (Fig.~\ref{fig:compared_spectra}(c)) differ from those reported by~\cite{Tarasenko} for acrylic samples excited using an electron beam and a KrCl excilamp. This could be due to differences in the excitation source, or the creation of colour centres~\cite{clough_color_1996,elgul_samad_production_2010}, or properties of the material itself such as the nature of additives and impurities.  
To the best of our knowledge, the  impurities mainly investigated in DEAP-type acrylic are radioactive ones from the uranium and thorium chains~\cite{nantais_radiopurity_2014}. 
In our data (Fig.~\ref{fig:compared_spectra}(e)), at lower temperatures, when the longer-wavelength peaks are most visible for the AVA2 sample, the wavelengths of those peaks are similar to the position of the spectral peaks of several compounds from the hydroxyphenyl benzotriazol (HBzT) class~\cite{fluegge_probing_2007}, including 2-(2-hydroxy-5-methylphenyl)benzotriazole (Tinuvin P), which can serve as a representative example~\cite{BEAVAN1974925}.
This suggests that if a similar compound is an additive in the AV acrylic, it could account for part of the luminescence observed in the sample. However, the room-temperature similarity with the DEAP-LG and SNO-AV samples implies there is some intrinsic photoluminescent mechanism as well, or a common photoluminescent additive.
Indeed, low-temperature fluorescence has been observed for certain HBzT compounds at $\sim 395$~nm~\cite{fluegge_probing_2007,KIRKBRIGHT1970237}. The overall spectral behavior can be explained as follows: at room temperature,  the PMMA dominates, with a broad, asymmetric peak around $410$~nm.  As the sample is cooled, the additives come into play, augmenting the short-wavelength structure and pushing it to slightly shorter wavelengths, and contributing the long wavelength features.


\subsection{\label{sec:normalized_pulses}Normalized Pulse Shapes}

To investigate the time structure of the TPB1 and AVA1 fluorescence pulses, in addition to the response with the samples in place, it is important to know the instrument response.  For this, we took data using the same setup but with neither sample nor broad bandpass filter.
The filter was removed because the LED emits light at 285~nm, and the broad bandpass has a lower wavelength limit of 375~nm, so it should block most of the light from the LED. 
The LED, AVA1 and TPB1 all produce different amounts of light in the detector so average pulses were normalized to unit area. 

\begin{figure}[H]
    \centering
    \includegraphics[width=0.7\textwidth]{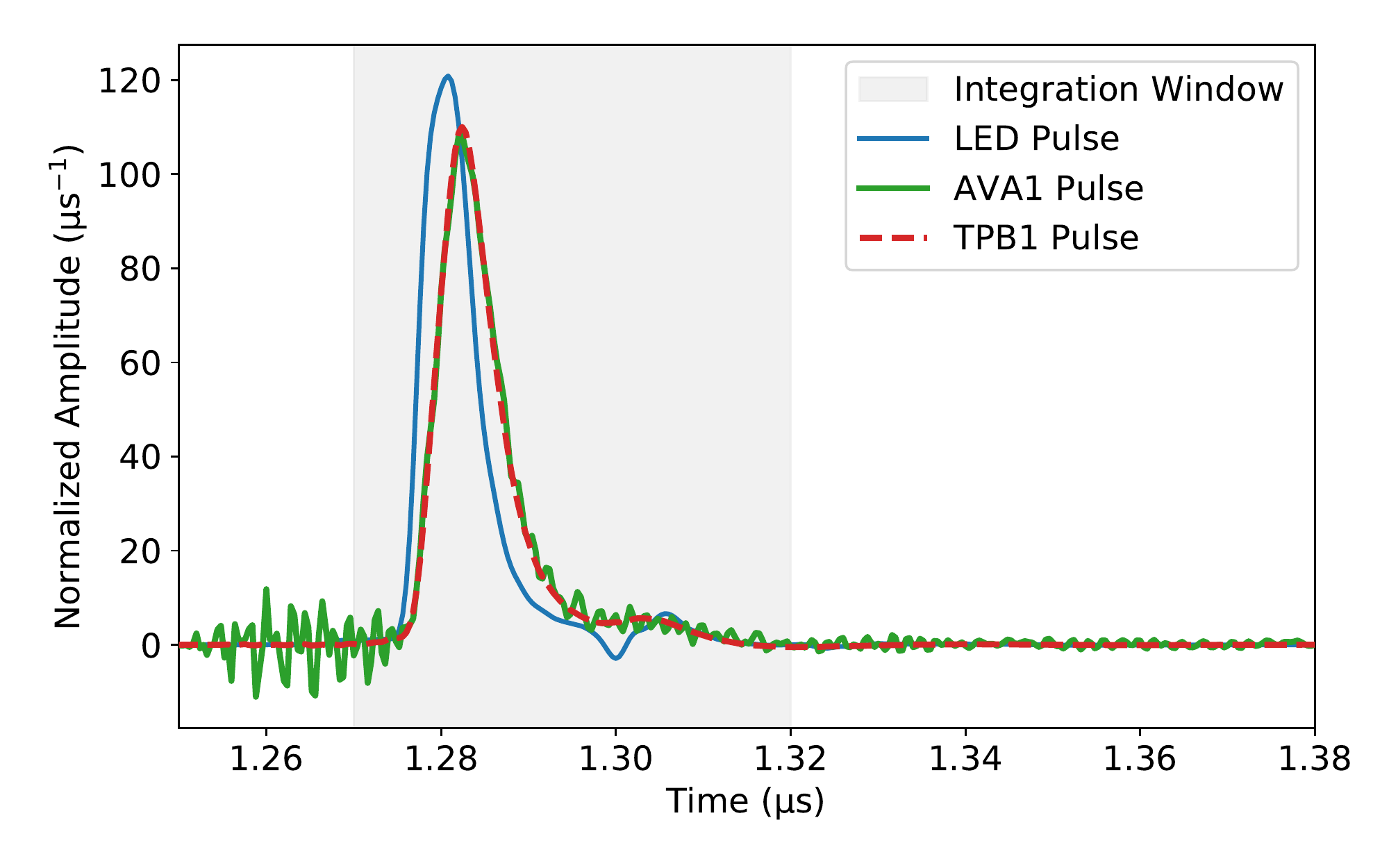}
    \caption{\label{fig:normalized_pulses}Average pulses from the LED directly shining light onto the PMT and from the AVA1 and TPB1 fluorescent light at 300~K. The shaded region indicates the 50~ns integration region used to calculate the light yield for AVA1 and TPB1. The response of the AVA1 and TPB1 are  similar to within the noise.}
\end{figure}

Fig.~\ref{fig:normalized_pulses} shows the comparison of the average pulses for the LED, AVA1, and TPB1 at 300~K, normalized to unit area. The shaded region indicates the 50~ns integration region that is used for the light yield analysis to help visualize where the integration window is relative to the fluorescence pulses. 
The TPB1 and AVA1 pulse shapes look  identical to within the noise, and are similar to that of the LED. This suggests that the pulse shape observed for both samples is dominated by the instrument response, which is a combination of the LED pulse shape, the time response of the PMT and readout electronics. It also suggests that the main time constants of TPB1 and AVA1 are shorter than a few nanoseconds. 
This is consistent with observation of a $\sim$~ns response of acrylic under excitation by a subnanosecond electron beam~\cite{Bolotovskii_2009}.
Spectral information (Sec.~\ref{sec:spectra}) suggests the presence of UV-absorber, possibly from the HBzT class,  in the AVA.  For such compounds, fluorescence and phosphorescence has been reported at low temperatures, the latter with a time constant of $\sim 1$~s~\cite{fluegge_probing_2007,KIRKBRIGHT1970237}. Our spectrum implies the amount of phosphorescence light is less than or equal to the amount of fluorescence light.  The latter is  less than a photoelectron per event on average in our time-resolved measurements (Sec.~\ref{sec:AVA_LY}).  As the phosphorescence photoelectrons are spread out over nearly a second, they would be masked by the few hundred Hertz of dark counts in our PMT, making phosphorescence unobservable in our pulse shape.

\subsection{\label{sec:TPB_LY}TPB Light Yield}

\begin{figure}[H]
    \centering
    \includegraphics[width=0.8\textwidth]{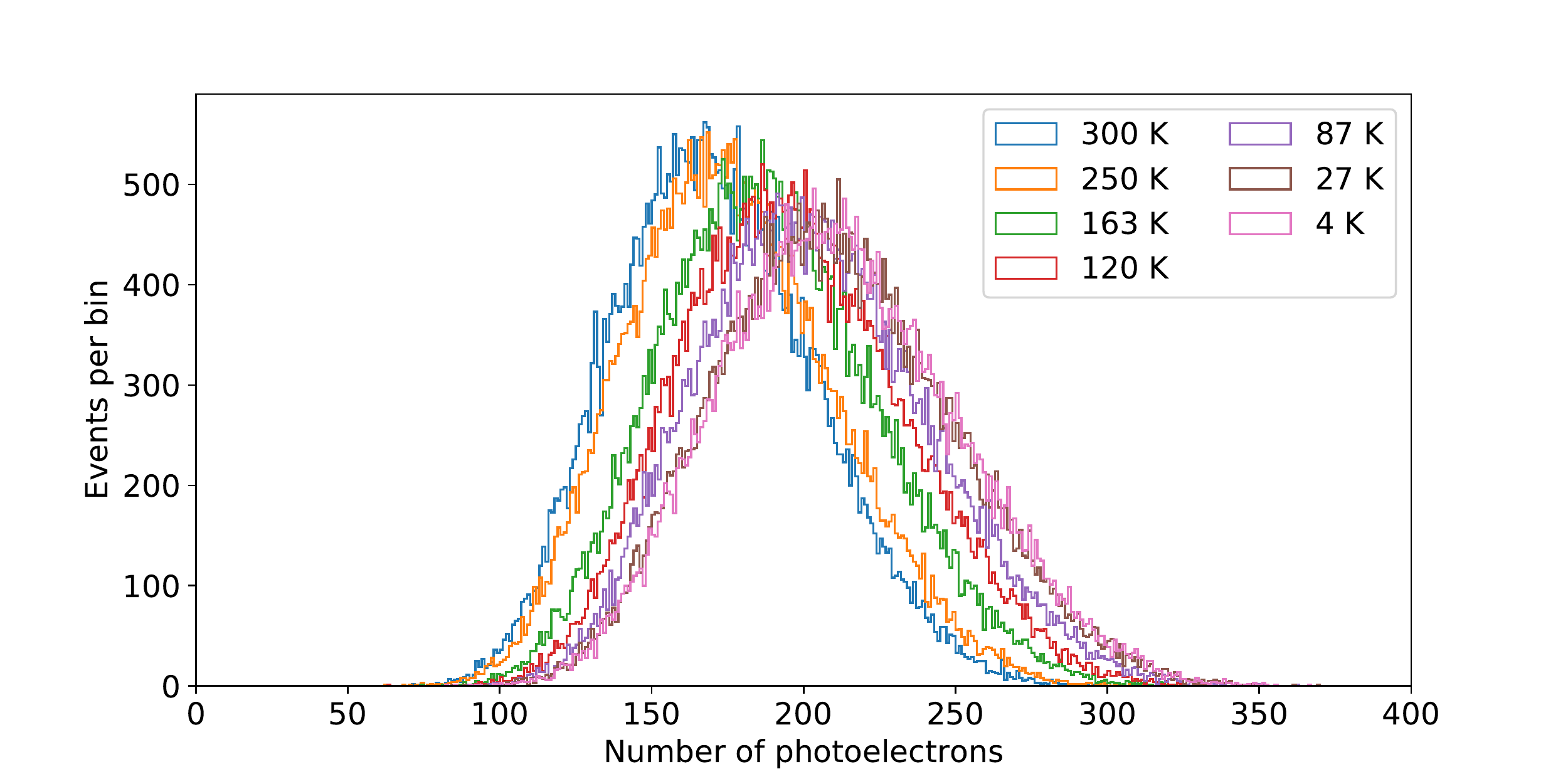}
    \caption{Integral distributions of TPB1 luminescence in terms of the number of photoelectrons at different temperatures.  Bin size is one photoelectrons.}
    \label{fig:TPB_distr}
\end{figure}

As explained in Sec.~\ref{sec:PulseAnalysis}, to determine the light yield, individual waveforms are integrated.
The integral distribution was fit with a skew normal function and the mean and error on the mean are used in the calculation of the light yield. Fig.~\ref{fig:TPB_distr} shows the integral distribution of TPB1 events at multiple temperatures in the number of photoelectrons. 
The systematic uncertainty for the TPB light yield was found to be 0.56~photoelectrons  stemming from the choice of integration window.

\begin{figure}[H]
    \centering
    \includegraphics[width=0.8\textwidth]{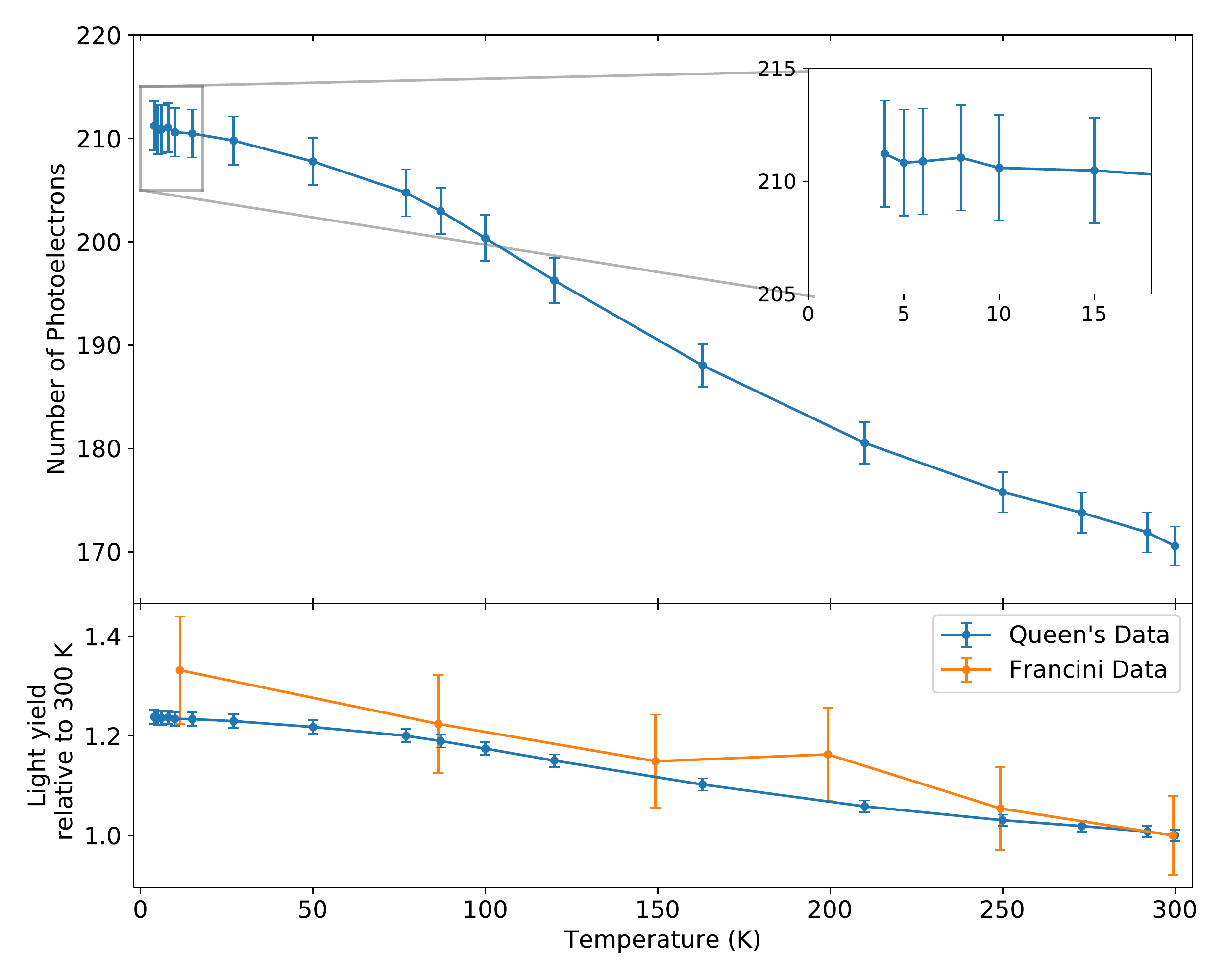}
    \caption{Top: TPB1 detected light yield as a function of temperature given in terms of the number of photoelectrons. Bottom: TPB1 detected light yield normalized to 300~K compared to results from Francini~\cite{Francini} also normalized to 300~K.  Evolution is consistent within errors.}
    \label{fig:TPB_LY}
\end{figure}

Fig.~\ref{fig:TPB_LY} shows that as the temperature decreases the light yield of the TPB1 sample increases. At 87~K there is a 19.0\% increase in light yield compared to 300~K, while at 4~K, the light yield has increased by 23.8\% relative to 300~K.  These numbers are consistent within errors with previous characterizations of TPB down to low temperatures~\cite{Francini}, and specifically at 87~K~\cite{araujo_rd_2021}.

\subsection{\label{sec:AVA_LY}AV Acrylic (AVA) Light Yield}

\begin{figure}[H]
    \centering
    \includegraphics[width=0.8\textwidth]{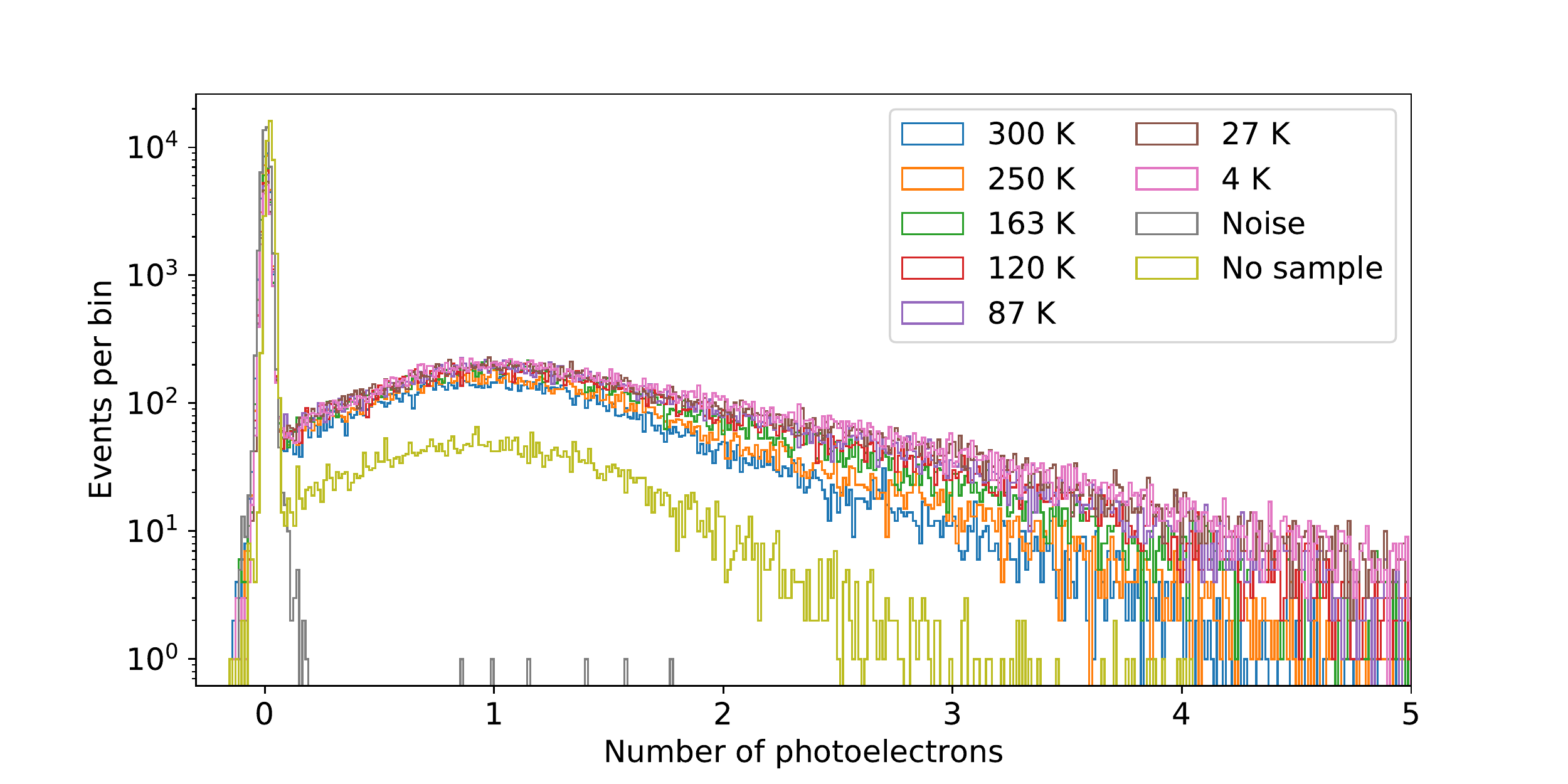}
    \caption{\label{fig:AVA_distributions}Integral distributions of AVA1 luminescence at different temperatures. The integral distribution for a noise run and a run when there was no sample in the detector are also shown.  The no-sample run averages less than a quarter of the photoelectrons per event than the runs with a sample (see text for details). Bin size is 0.13~photoelectrons.}
\end{figure}
Fig.~\ref{fig:AVA_distributions} shows the distribution of event integrals from the AVA1 sample at various temperatures.
The numbers of photoelectrons are quite low, so it is important to confirm that they come from fluorescence of the sample and are not some form of background.  Fig.~\ref{fig:AVA_distributions} also includes data taken with no sample.  Photoelectrons are observed, but at a significantly lower level ($0.120 \pm 0.004 $~SPE/evt) than the lowest level when the sample is present ($0.465 \pm 0.007$~SPE/evt).  This low background may come from residual LED light making it through the various filters, or possibly from fluorescence of cryostat material such as low-level fluorescence of glasses and filters.  Regardless of the origin of the light, this no-sample measurement probably overestimates the amount of background light when the sample is in place, since in the no-sample measurement the LED has a direct line of sight to the PMT, whereas the sample may act as a baffle.  For this reason, we consider the measured light the actual fluorescence of the sample, as opposed to an upper limit on it.
Note that if we integrate all noise events in the same 50~ns window as the LED pulse data, we get a zero-centred distribution, known as the pedestal, also shown on Fig.~\ref{fig:AVA_distributions}, and much narrower than the SPE integrals.

\begin{figure}[H]
    \centering
    \includegraphics[width=0.7\textwidth]{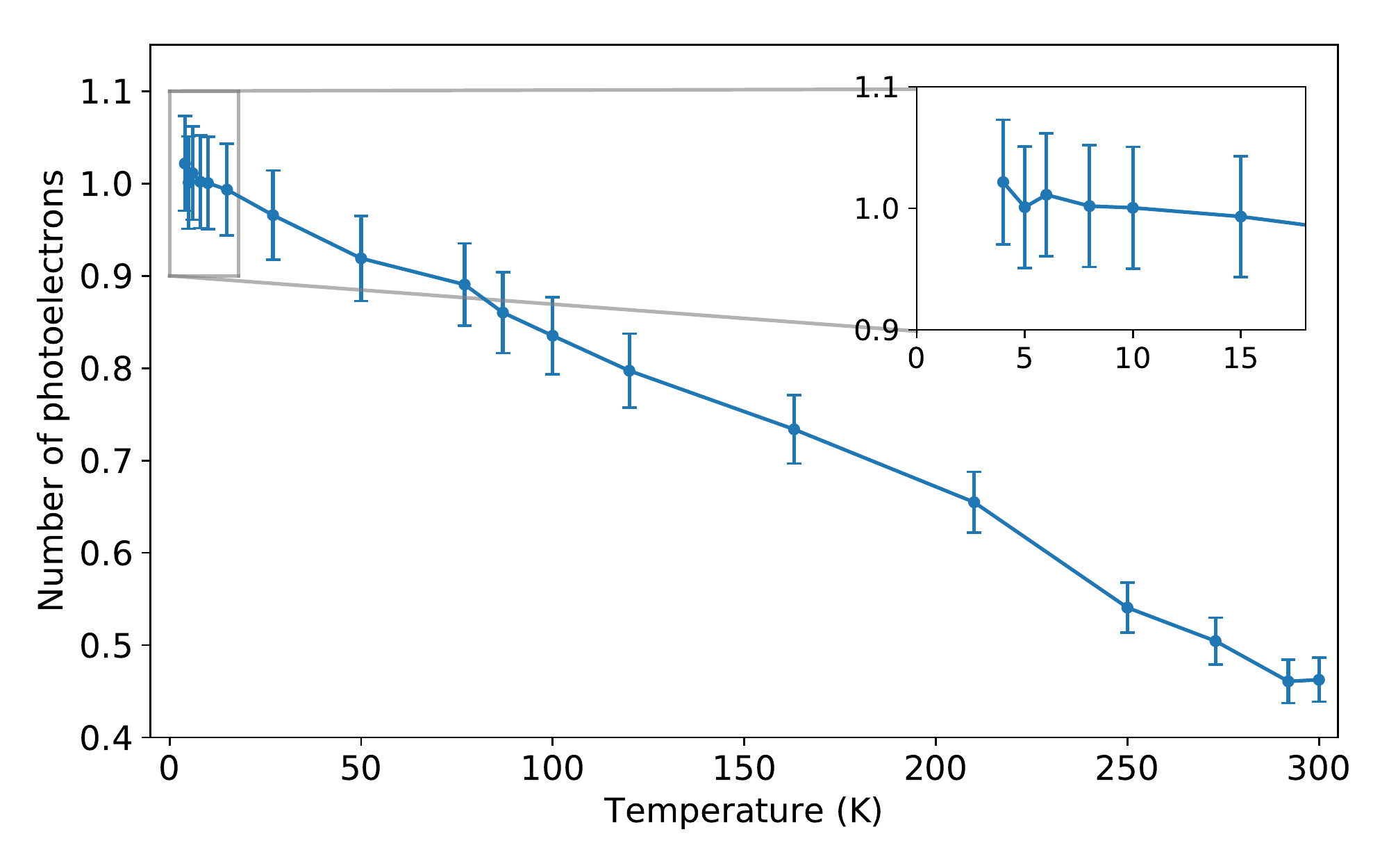}
    \caption{AVA1 detected light yield in number of photoelectrons at multiple temperatures.}
    \label{fig:AVA_LY}
\end{figure}

From the distributions of photoelectrons at each temperature  in Fig.~\ref{fig:AVA_distributions}, we obtain the evolution of the light yield as a function of temperature in Fig.~\ref{fig:AVA_LY}.  The trend is similar to TPB1 as the light yield increases with decreasing temperature. The overall fluorescence light yield of AVA1 is quite low; at most temperatures less than 1 photoelectron was observed on average per event. Relative to the light yield at 300~K, at 87~K the light yield increased by $\sim 85$\% while at 4~K it had increased by $\sim 120$\%. The systematic error on the light yield was calculated by the same method used in the analysis of the TPB1 sample, i.e. looking at different integration windows,  and was set to 0.04~SPE for these measurements.

\begin{figure}[H]
    \centering
    \includegraphics[width=0.7\textwidth]{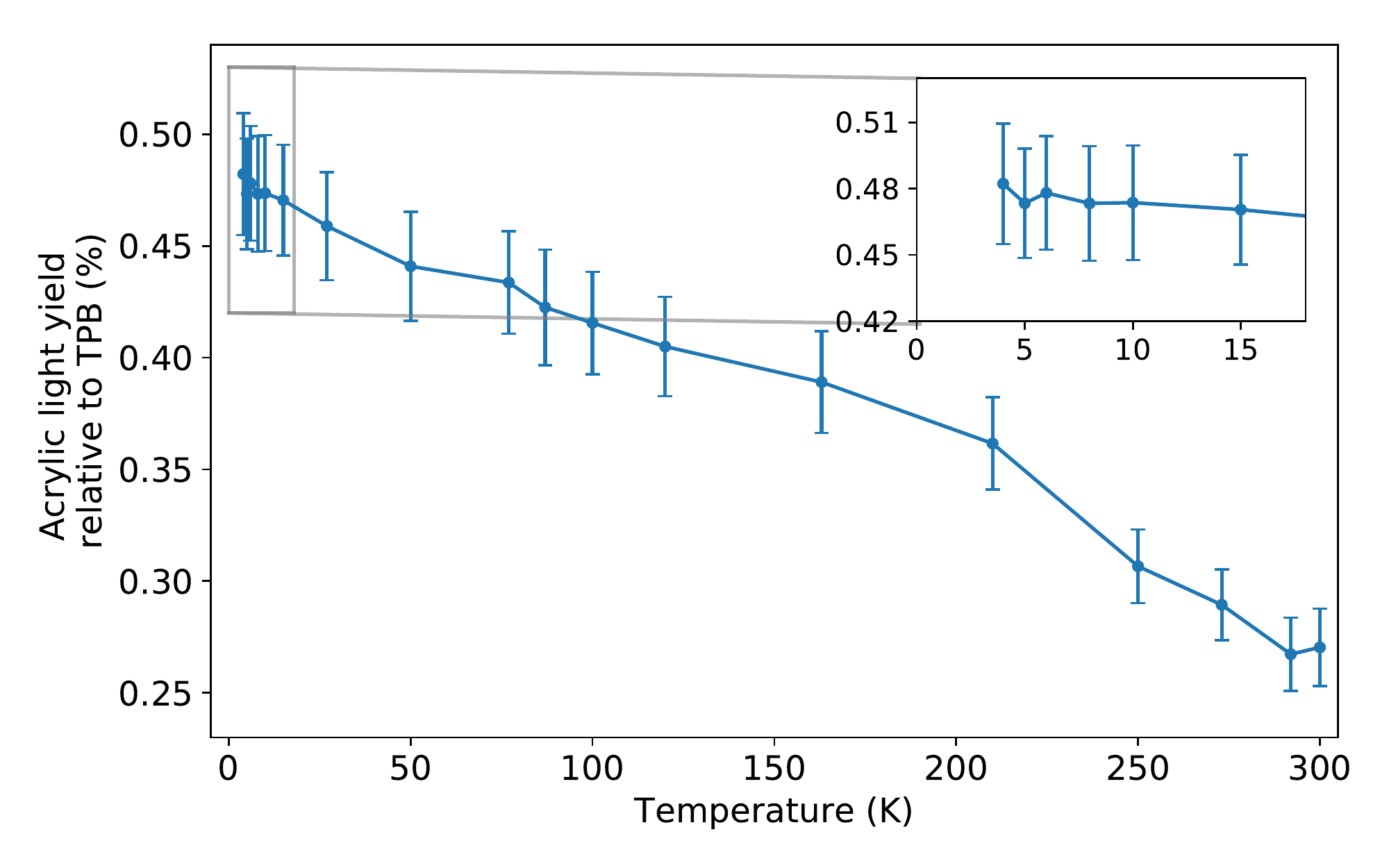}
    \caption{AVA1 light yield relative to TPB1 at multiple temperatures.}
    \label{fig:AVA_RLY}
\end{figure}

It is possible to calculate a relative light yield between AVA1 and TPB1 by applying corrections for the LED voltage (Sec.~\ref{sec:LY} and \ref{sec:CorFact}),  and accounting for the differing vertical ranges used on the digitizer. Fig.~\ref{fig:AVA_RLY} shows this relative light yield at different temperatures. The relative light yield varies from approximately 0.3\% at 300~K to 0.5\% at 4~K.  A previously published work, with similar samples, but different excitation wavelengths and methods, sets an upper limit of 0.2\% at 300~K~\cite{RPT_LY}. It also shows the response depends on various factors, including the excitation wavelength in particular.

\section{Conclusion}

We studied the fluorescence of acrylic and TPB-coated acrylic samples in a cryostat using an excitation wavelength of 285~nm. The fluorescence spectra and pulseshapes were obtained in a range of temperatures from 300~K down to 4~K. The TPB spectra are consistent with previous work~\cite{Francini,LIDINE} and well-resolved at low temperatures, where an increased light yield and reduced thermal broadening reveal substructure including peaks at $\sim 400$~nm and $\sim 425$~nm.
The acrylic spectra show one main peak with a maximum at $\sim 395$~nm and additional longer-wavelength features at low temperatures. 
The long-wavelength features are attributed to phosphorescence of UV-absorber in the acrylic.  Short-wavelength similarities between spectra of different acrylic samples suggest that in addition to the contribution from additives, part of the fluorescence is related to the acrylic itself, though we can't exclude that an additive common to all the samples plays a role.

The light yields of both acrylic and TPB increase as the temperature falls. The TPB light yield  shows a similar evolution with temperature as reported in~\cite{Francini}. The light yield of acrylic relative to TPB also increases with decreasing sample temperature up to a maximum of 0.5\%. At 300~K the acrylic relative light yield reaches 0.3\%. This is close to the 0.2\% reported in earlier work~\cite{RPT_LY} despite that work using a shorter excitation wavelength and different samples in a completely different experimental setup. 
The dominant time constants of acrylic and TPB are shorter than the $\sim 10$~ns instrument response. 

Understanding the fluorescence of acrylic is important for rare-event searches  using that material as a detector component because it can contribute to the background, for instance when superluminal charged particles produce Cherenkov light in the acrylic. 
As  impurities and additives may contribute to the fluorescence, a good knowledge of the fluorescence properties of the specific acrylic used in a given experiment is therefore needed to account for these background events.

\section{Acknowledgements}

Funding in Canada has been provided by NSERC through SAPPJ grants, by CFI-LOF and ORF-SIF, and by the McDonald Institute.
Prof.~M.~Ku\'zniak is supported by the International Research Agenda Programme AstroCeNT (MAB\allowbreak/2018\allowbreak/7) funded by the Foundation for Polish Science (FNP) from the European Regional Development Fund. AstroCeNT and Technical University of Munich cooperation is supported from the EU’s Horizon 2020 research and innovation program under grant agreement No 962480 (DarkWave).
Prof.~M.~Chen and the SNO+ group at Queen's kindly allowed access to their Photon Technology International QuantaMaster fluorescence spectrophotometer, and provided the  acrylic samples from the SNO experiment.

\bibliographystyle{elsarticle-num}
\bibliography{ms.bib}

\clearpage

\appendix

\section{\label{sec:CorFact}Correction factor}

At each temperature $T$, we measure a number of photoelectrons from AVA1 at a 13.7~V LED voltage ($n_{AVA1}(T; 13.7)$), and a number of photoelectrons from TPB1 at a lower 13.4~V LED voltage ($n_{TPB}(T; 13.4)$).
The LED voltage is lower for TPB1 because at the AVA1 voltage (13.7~V), pulses  saturate the largest vertical range of the digitizer at low temperatures. To compare the relative light yields,  we want  the ratio for the same excitation (ie LED voltage): $\frac{n_{AVA1}(T; 13.7)}{ n_{TPB1}(T; 13.7)}$.  
These are related to the measurements by:

$$\underbrace{\frac{n_{AVA1}(T; 13.7)}{ n_{TPB1}(T; 13.7)}}_{\text{relative light yield}} = \underbrace{ \frac{n_{AVA1}(T; 13.7)}{ n_{TPB1}(T; 13.4)}}_{\text{measurement}} \times \underbrace{ \frac{n_{TPB1}(T; 13.4)}{n_{TPB1}(T; 13.7)}}_{\text{correction}}$$
We measure the term $ \frac{n_{TPB}(T; 13.4)}{n_{TPB1}(T; 13.7)}$ (correction factor) at three temperatures close to room temperature, and assume it does not depend on temperature.  
Numerical values of the correction factor are provided in Tab.~\ref{tab:tpb_cor_factor}.
\begin{table}[h]
\centering
\begin{tabular}{|llll}
\hline
\multicolumn{1}{|l}{Temperature} & Light Yield @ 13.7~V & Light Yield @ 13.4~V & \multicolumn{1}{l|}{Correction} \\
\multicolumn{1}{|l}{(K)} & (photoelectrons) & (photoelectrons) & \multicolumn{1}{l|}{Factor} \\ \hline \hline

\multicolumn{1}{|l}{300} & 171.2 $\pm$ 0.5 & 67.0 $\pm$ 0.3 & \multicolumn{1}{l|}{2.556 $\pm$ 0.009} \\ \hline
\multicolumn{1}{|l}{292} & 171.6 $\pm$ 0.5 & 67.5 $\pm$ 0.3 & \multicolumn{1}{l|}{2.542 $\pm$ 0.009} \\ \hline
\multicolumn{1}{|l}{273} & 173.4 $\pm$ 0.5 & 68.2 $\pm$ 0.3 & \multicolumn{1}{l|}{2.542 $\pm$ 0.009} \\ \hline
\multicolumn{1}{l}{} & & \textbf{Average } & 2.547 $\pm$ 0.005 \\

\end{tabular}

\caption{\label{tab:tpb_cor_factor}TPB1 correction factor calculation based on the light yields at the lower LED voltage used for the whole run at 13.4~V compared to the higher voltage used for select temperatures at 13.7~V.}
\end{table}

\end{document}